\documentclass[onecolumn,amsmath,showpacs,nofootinbib,11pt]{revtex4}
\usepackage{graphicx}
\usepackage{dcolumn}
\usepackage{bm}
\begin{document}
\newcommand{\hs}{\hspace*{0.5cm}}
\newcommand{\vs}{\vspace*{0.5cm}}
\newcommand{\be}{\begin{equation}}
\newcommand{\ee}{\end{equation}}
\newcommand{\bea}{\begin{eqnarray}}
\newcommand{\eea}{\end{eqnarray}}
\newcommand{\ben}{\begin{enumerate}}
\newcommand{\een}{\end{enumerate}}
\newcommand{\bde}{\begin{widetext}}
\newcommand{\ede}{\end{widetext}}
\newcommand{\nn}{\nonumber}
\newcommand{\crn}{\nonumber \\}
\newcommand{\Tr}{\mathrm{Tr}}
\newcommand{\non}{\nonumber}
\newcommand{\noi}{\noindent}
\newcommand{\al}{\alpha}
\newcommand{\la}{\lambda}
\newcommand{\bet}{\beta}
\newcommand{\ga}{\gamma}
\newcommand{\va}{\varphi}
\newcommand{\om}{\omega}
\newcommand{\pa}{\partial}
\newcommand{\+}{\dagger}
\newcommand{\fr}{\frac}
\newcommand{\bc}{\begin{center}}
\newcommand{\ec}{\end{center}}
\newcommand{\Ga}{\Gamma}
\newcommand{\de}{\delta}
\newcommand{\De}{\Delta}
\newcommand{\ep}{\epsilon}
\newcommand{\varep}{\varepsilon}
\newcommand{\ka}{\kappa}
\newcommand{\La}{\Lambda}
\newcommand{\si}{\sigma}
\newcommand{\Si}{\Sigma}
\newcommand{\ta}{\tau}
\newcommand{\up}{\upsilon}
\newcommand{\Up}{\Upsilon}
\newcommand{\ze}{\zeta}
\newcommand{\ps}{\psi}
\newcommand{\Ps}{\Psi}
\newcommand{\ph}{\phi}
\newcommand{\vph}{\varphi}
\newcommand{\Ph}{\Phi}
\newcommand{\Om}{\Omega}

\title{The economical 3-3-1 model revisited}   

\author{P. V. Dong}\email{pvdong@iop.vast.ac.vn}\affiliation{Institute of Physics, Vietnam Academy of Science and Technology, 10 Dao Tan, Ba Dinh, Hanoi, Vietnam}

\author{D. Q. Phong}\email{tphongvltckt@yahoo.com}\affiliation{Department of Physics, Hanoi Architectural University, 10 Nguyen Trai, Ha Dong, Hanoi, Vietnam} 

\author{D. V. Soa}\email{dvsoa@assoc.iop.vast.ac.vn}\affiliation{Department of Physics, Hanoi Metropolitan University, 98 Duong Quang Ham, Cau Giay, Hanoi, Vietnam}

\author{N. C. Thao}\email{ncthao@grad.iop.vast.ac.vn}\affiliation{Graduate University of Science and Technology, Vietnam Academy of Science and Technology, 18 Hoang Quoc Viet, Cau Giay, Hanoi, Vietnam}

\begin{abstract}

We show that the economical 3-3-1 model poses a very high new physics scale of the order of 1000~TeV due to the constraint on the flavor-changing neutral current. The implications of the model for neutrino masses, inflation, leptogenesis, and superheavy dark matter are newly recognized. Alternatively, we modify the model by rearranging the third quark generation differently from the first two quark generations, as well as changing the scalar sector. The resultant model now predicts a consistent new physics at TeV scale unlike the previous case and may be fully probed at the current colliders. Particularly, due to the minimal particle contents, the models under consideration manifestly accommodate dark matter candidates and neutrino masses, with novel and distinct production mechanisms. The large flavor-changing neutral currents that come from the ordinary and exotic quark mixings can be avoided due to the approximate $B-L$ symmetry.          

\end{abstract}

\date{\today}

\maketitle

\section{\label{introduction} Introduction}

There are now the certain experimental evidences that require new physics beyond the standard model. They mainly include neutrino oscillations, the baryon asymmetry of the universe, dark matter, and the cosmic inflation \cite{pdg}. Traditional proposals such as supersymmetry, extradimension, and grand unification can solve only some of the questions separately and obey several issues on both the theoretical and experimental sides \cite{pdg}. In this work, we show that the model based upon the gauge symmetry $SU(3)_C\otimes SU(3)_L\otimes U(1)_X$ (3-3-1) \cite{331m,331r} may be an intriguing choice for the new physics, besides its ability to provide common answers to most of these puzzles.         
        
Indeed, the new weak isospin group $SU(3)_L$ that is directly extended from the $SU(2)_L$ symmetry of the standard model is well-motivated due to its ability to obtain the number of generations to match that of fundamental colors by the $[SU(3)_L]^3$ anomaly cancelation. However, the electric charge $Q$ neither commutes nor closes algebraically with $SU(3)_L$, analogous to the standard model. Hence, a new Abelian group $U(1)_X$ is derived as a result to close those symmetries by the gauge group $SU(3)_L\otimes U(1)_X$, which includes $Q=T_3+\beta T_8+X$ as a residual charge, where $T_n\ (n=1,2,3,...,8)$ and $X$ denote the $SU(3)_L$ and $U(1)_X$ charges, respectively (cf. \cite{d}). Imposing the color group $SU(3)_C$, one has the complete gauge symmetry $SU(3)_C\otimes SU(3)_L\otimes U(1)_X$, as mentioned. The nontrivial commutations for $Q$ are $\left[Q, T_1\pm i T_2\right] = \pm (T_1\pm i T_2)$, $\left[Q, T_4\pm i T_5\right] = \mp q (T_4\pm i T_5)$, and $\left[Q, T_6\pm i T_7\right] = \mp (1+q) (T_6\pm i T_7)$, where the last two relations define the electric charges of new particles in representations via a basic electric charge $q\equiv -(1+\sqrt{3}\beta)/2$. Let us stress that $\beta$ (thus $q$) is arbitrary on the theoretical ground and is independent of all anomalies \footnote{The introduction of the embedding coefficient $\beta$ was given early in \cite{331beta1,331beta2}.}.    

The general 3-3-1 model including the original versions \cite{331m,331r} have plenty of fields, three scalar triplets, $\eta=(\eta^0_1,\eta^-_2,\eta^q_3)$, $\rho=(\rho^+_1,\rho^0_2,\rho^{q+1}_3)$, $\chi=(\chi^{-q}_1,\chi^{-q-1}_2,\chi^0_3)$, with/without one scalar sextet, $S=(S^0_{11},S^-_{12},S^q_{13},S^{--}_{22},S^{q-1}_{23},S^{2q}_{33})$, and more exotic fermions, as the choice for mass generation and/or anomaly cancelation. However, those particle contents are complicated, preventing the models' predictability. Therefore, we would like to search for some calculable 3-3-1 model that contains a minimal content of fermions and scalars. Following this approach, the first one was the economical 3-3-1 model working only with two scalar triplets $\rho,\chi$ \cite{e1,e2}, extracted from the 3-3-1 model with right-handed neutrinos, for $\beta=-1/\sqrt{3}$ \cite{331r}. The new physics implications as well as the supersymmetric extension were extensively investigated in \cite{e3,e4,e5,e6,e7,e8,e9,e10,e11,ed1,ed2,ed3,ed4,e12}. The second one was first introduced as the reduced 3-3-1 model working with $\rho,\chi$ \cite{331rm}, deduced from the minimal 3-3-1 model, for $\beta=-\sqrt{3}$ \cite{331m}. However, such version was encountered with the problems of the $\rho$-parameter bound, the flavor-changing neutral current (FCNC) constraint, and the Landau pole limit (cf. \cite{ds}). The realistic theory for the second approach that overcomes such issues was finally achieved, called the simple 3-3-1 model \cite{s331}, which works with $\eta,\chi$ and has the phenomenological aspects extensively studied in \cite{s331dm,s331pheno}. 

Although not presenting a low Landau pole, the economical 3-3-1 model may encounter the other bounds similarly to the reduced 3-3-1 model, which lead to the interesting consequences to be examined in this work. To be concrete, we reconsider the new physics scale of the economical 3-3-1 model due to the FCNC constraint. We show that there are such two distinct versions of it. The first version is just the original model, but it works surprisingly with a new physics scale beyond 1000 TeV. As a viable high energy regime, this case provides naturally a seesaw mechanism, inflationary scenario, leptogenesis, and superheavy dark matter. These features are in sharp contrast with the previous interpretations in the old model \cite{e1,e2,e3,e4,e5,e6,e7,e8,e9,e10,e11,ed1,ed2,ed3,ed4}. The second version arises once the fermion and scalar contents are appropriately recast (i.e., changed), yielding a low new physics scale just beyond 1 TeV. This setup provides new physics consequences, such as the neutrino mass mechanism, new fermions, new Higgs and gauge bosons, and the weakly-interacting massive particle (WIMP), which may be fully probed at the current colliders.    

For the purpose, in Sec. \ref{fcnc}, we consider the 3-3-1 model with arbitrary $\beta$ and extract the bound for the 3-3-1 breaking scale due to the FCNC constraints. As we see, this bound of the new physics scale depends only on the arrangement of quark representations. In Sec. \ref{t12e331}, applying the above result to the economical 3-3-1 model, two folds for the model are derived, and their corresponding consequences are discussed. Let us remind the reader that these variants of the economical 3-3-1 model are not limited by a Landau pole, because this pole is actually higher than the Planck scale, as proven in \cite{landaupole}. Finally, we conclude this work in Sec. \ref{con}.  

\section{\label{fcnc} FCNCs}

The 3-3-1 model with arbitrary $\beta$ is given by the electric charge operator,
\be Q=T_3+\beta T_8+X, \ee as mentioned. Therefore, the hypercharge is obtained as $Y=\beta T_8+X$. Furthermore, the fundamental representations of $SU(3)_L$ can be decomposed as $3=2\oplus 1$ and $3^*=2^*\oplus 1$ under $SU(2)_L$. Here, the antidoublet takes the form $(f_2,-f_1)$, given that $(f_1,f_2)$ is a doublet. Hence, all the left-handed fermion doublets will be enlarged to 3 or $3^*$, while the right-handed fermion singlets are retained as $SU(3)_L$ singlets, or suitably combined with the above fermion doublets. The $SU(3)_L$ anomaly cancelation requires the number of 3 to be equal that of $3^*$, where the color number is appropriately counted. Thus, assuming that the first quark generation transforms under $SU(3)_L$ differently from the last two quark generations, the fermion content is achieved as 
\bea && \psi_{aL} =  
\left(\begin{array}{c}
\nu_{aL}\\
e_{aL}\\
k_{aL}\end{array}\right)  \sim \left(1,3,\fr{-1+q}{3}\right),\\
&& Q_{1 L}= 
\left(\begin{array}{c}
u_{1L}\\
d_{1L}\\
j_{1L}\end{array}\right)   \sim \left(3,3,\fr{1+q}{3}\right),\\ 
&& Q_{\al L} = 
\left(\begin{array}{c}
d_{\al L}\\
-u_{\al L}\\
j_{\al L} \end{array}\right)  \sim \left(3,3^*,-\fr{q}{3}\right),\\
&& e_{aR}\sim (1,1,-1),\hs k_{aR}\sim (1,1,q),\hs \nu_{aR}\sim (1,1,0),\\ 
&&u_{a R}\sim \left(3,1,\fr 2 3\right),\hs d_{aR}\sim \left(3,1,-\fr 1 3\right), \\
&& j_{1R}\sim \left(3,1,\fr 2 3 +q\right),\hs j_{\al R}\sim \left(3,1,-\fr 1 3 -q\right), \eea
where $a=1,2,3$ and $\al=2,3$ are generation indices. The numbers in parentheses denote representations based upon the $SU(3)_C$, $SU(3)_L$, and $U(1)_X$ groups, respectively. The above fermion pattern is free from all the other anomalies too.

The new fermions $k_a, j_a$ have been included to complete the representations, where their electric charges are related to the basic electric charge $q=-(1+\sqrt{3}\beta)/2$ through $Q(k_a)=q$, $Q(j_1)=q+2/3$, and $Q(j_\al)=-q-1/3$, as stated. The right-handed neutrinos $\nu_{aR}$ are sterile, i.e. gauge singlets, which may be imposed or not. This feature also applies for $k_{aR}$ if $q=0$. Moreover, two minimal 3-3-1 versions have traditionally been studied, provided that $\nu_{aR}$ and $k_{aR}$ are omitted, while $k_{aL}$ are replaced by either $(e_{aR})^c$ or $(N_{aR})^c$, respectively \cite{331m,331r}. $N_{aR}$ are some neutral fermions like the right-handed neutrinos. Such ingredient does not work for quarks, i.e. $j_{aL}$ cannot be substituted by ordinary right-handed quarks, because $SU(3)_C$, $SU(3)_L$, and spacetime symmetry commute. Thus, the inclusion of $j_a$ is necessary. On the other hand, if the second or third quark generation is arranged differently from the remaining quark generations, by contrast, the index $\al$ will take values, $\al=1,3$ or $\al=1,2$, respectively. Let us remind the reader that the following results can be generalized to all cases of the quark arrangements. 

Typically, the scalar content includes
\bea \eta &=&
\left(
\begin{array}{l}
\eta^{0}_1\\
\eta^{-}_2\\
\eta^{q}_3
\end{array}\right)\sim \left(1,3,\fr{q-1}{3}\right),\\
\rho &=&
\left(
\begin{array}{l}
\rho^{+}_1\\
\rho^{0}_2\\
\rho^{q+1}_3
\end{array}\right)\sim \left(1,3,\fr{q+2}{3}\right),\\
\chi &=&
\left(
\begin{array}{l}
\chi^{-q}_1\\
\chi^{-q-1}_2\\
\chi^{0}_3
\end{array}\right)\sim\left(1,3,-\fr{2q+1}{3}\right).\eea Here, the superscripts stand for electric charge values, while the subscripts indicate component fields under $SU(3)_L$. The scalars have such quantum numbers since they couple a left-handed fermion to a corresponding right-handed fermion to perform the relevant Yukawa Lagrangian. 

When the scalar triplet, $\chi$, develops a vacuum expectation value (VEV), $\langle \chi\rangle=\fr{1}{\sqrt{2}}(0\ 0\ w)^T$, it breaks the 3-3-1 symmetry down to the standard model and generates the masses for the new particles. Hence, this scalar must be introduced. The scalar triplets, $\eta$ and $\rho$, which have VEVs, $\langle \eta\rangle =\fr{1}{\sqrt{2}}(u\ 0\ 0)^T$ and $\langle \rho \rangle =\fr{1}{\sqrt{2}}(0\ v\ 0)^T$, break the standard model symmetry down to $SU(3)_C\otimes U(1)_Q$ and give the masses for the ordinary particles. Note that the other components of the scalar triplets may have a nonzero VEV, if they are electrically neutral. But such VEV can be strongly suppressed (cf., for instance, \cite{dhqt}). The minimal 3-3-1 model and the 3-3-1 model with right-handed neutrinos work with the above three scalar triplets. Even, they impose additional scalar multiplets, e.g. the scalar sextets. However, the simple 3-3-1 model and the economical 3-3-1 model work only with two scalar triplets, ($\chi$, $\eta$) and ($\chi$, $\rho$), respectively. 

In Ref. \cite{dl}, we have pointed out that due to $U(1)_Q$ invariance, the gauge boson spectrum and the gauge coupling matching are always determined for the 3-3-1 model with arbitrary $\beta$ and scalar sector, which will be used for the following analysis.     

Because the quark generations are not universal under the $SU(3)_L\otimes U(1)_X$ gauge symmetry, there are FCNCs. Indeed, the neutral current takes the form:  
\bea \mathcal{L} &\supset& \bar{F}i\ga^\mu D_\mu F\crn
&\supset& -g\bar{F}\ga^\mu[T_3 A_{3\mu} + T_8 A_{8\mu}+t_X(Q-T_3-\beta T_8)B_\mu]F,\eea where $F$ runs over all fermion multiplets, and the covariant derivative $D_\mu=\pa_\mu+ig_s t_n G_{n\mu}+ig T_n A_{n\mu}+ig_X X B_\mu$ contains, by definition, the coupling constants $(g_s,g,g_X)$, the generators $(t_n,T_n,X)$, and the gauge bosons $(G_n, A_n,B)$ of the $SU(3)_C$, $SU(3)_L$, and $U(1)_X$ groups, respectively. We have also used $X=Q-T_3-\beta T_8$ and $t_X\equiv g_X/g=t_W/\sqrt{1-\beta^2 t^2_W}$ \cite{dl}. The ordinary leptons and the new fermions do not flavor change, because the corresponding flavor groups that potentially mix (within each group), such as $\{\nu_{aL}\}$, $\{e_{aL}\}$, $\{e_{aR}\}$, $\{k_{aL}\}$, $\{k_{aR}\}$, $\{j_{\al L}\}$, and $\{j_{\al R}\}$, are respectively identical under the gauge charges. Simultaneously, the terms of $T_3$ and $Q$ do not leading to flavor changing, because all the mentioned flavor groups including the ordinary quarks $\{u_{aL}\}$, $\{u_{aR}\}$, $\{d_{aL}\}$, and $\{d_{aR}\}$ are respectively identical under these charges ($T_3,Q$) \footnote{For the 3-3-1 models without exotic charges (i.e., $q=0$ or $-1$), the ordinary quarks and the exotic quarks that have different weak isospins might mix, leading to large FCNCs associated with $Z$ boson, independent of the generation nonuniversality \cite{dhqt}. This effect might be more dangerous than the nonuniversal $Z'$ couplings, and is only suppressed if such mixing is small compared to the mixing of the ordinary quarks, as shown below. Alternatively, the FCNCs may be associated with the neutral scalars as discussed in \cite{fcnc3,fcnc1,fcnc2}.}. Thus, the FCNCs only couple the ordinary quarks to $T_8$, arising in part from 
\bea \mathcal{L}\supset -g\bar{q}_L\ga^\mu T_{8L}q_L (A_{8\mu}-\beta t_X B_\mu). \eea
Here we denote either $q=(u_1,u_2,u_3)$ for up quarks or $q=(d_1,d_2,d_3)$ for down quarks, and $T_{8L}=\fr{1}{2\sqrt{3}}\mathrm{diag}(1,-1,-1)$ summarizes the $T_8$ values of $q_{1L}$, $q_{2L}$, and $q_{3L}$, respectively. 

Changing to the mass basis, we have $q_{L,R}=V_{qL,qR}q'_{L,R}$, where $q'$ is either $q'=(u,c,t)$ or $q'=(d,s,b)$, and $V_{qL,qR}$ are the quark mixing matrices that diagonalize the corresponding mass matrices, $V^\dagger_{uL} M_u V_{uR}=\mathrm{diag}(m_u,m_c,m_t)$ and $V^\dagger_{dL} M_d V_{dR}=\mathrm{diag}(m_d,m_s,m_b)$. Further, the Cabibbo-Kobayashi-Maskawa (CKM) matrix takes the form, $V_{\mathrm{CKM}}=V_{uL}^\dagger V_{dL}$. With the aid of $A_{8\mu}- \beta t_X B_\mu =(1/\sqrt{1-\beta^2 t^2_W})Z'_\mu$ \cite{dl}, it follows  
\bea \mathcal{L} &\supset& -\fr{g}{\sqrt{1-\beta^2 t^2_W}}\bar{q}'_L\ga^\mu (V^\dagger_{qL}T_{8L} V_{qL})q'_L Z'_\mu, \crn
&\supset& -\fr{g}{\sqrt{3(1-\beta^2 t^2_W)}}\bar{q}'_{iL}\ga^\mu q'_{jL} (V^*_{qL})_{1i}(V_{qL})_{1j}Z'_\mu, \eea which causes tree-level FCNCs for $i\neq j$, where $i,j=1,2,3$ label respective physical quark states in $q'$. The new neutral gauge boson $Z'$ might mix with the standard model $Z=c_W A_3 - s_W(\beta t_W A_8+\sqrt{1-\beta^2 t^2_W} B)$ and the real part of new non-Hermitian gauge bosons, $V=A_4$ or $V=A_6$, for $q=0$ or $q=-1$, respectively. The contribution of $V$ to the FCNCs is negligible, which can be justified, using \cite{e2}. Therefore, we write $Z'=-s_\varphi Z_1 + c_\varphi Z_2$, where $Z_{1,2}$ are two physical neutral gauge bosons with masses 
\be m^2_{Z_1}\simeq \fr{g^2}{4c^2_W}(u^2+v^2),\hs m^2_{Z_2}\simeq \fr{g^2w^2}{3(1-\beta^2 t^2_W)},\ee
and the $Z$-$Z'$ mixing angle is 
\be t_{2\varphi}  \simeq  \dfrac{\sqrt{3(1-\beta^2t^2_W)}}{2c_W w^2}  \left[(1+\sqrt3\beta t^2_W)u^2 -(1-\sqrt3\beta t^2_W)v^2\right].\ee 

Substituting $Z'$ into the above FCNCs and integrating $Z_{1,2}$ out, we obtain the effective Lagrangian describing meson mixings, 
\be \mathcal{L}^{\mathrm{eff}}_{\mathrm{FCNC}}=\fr{g^2}{3(1-\beta^2 t^2_W)} (\bar{q}'_{iL}\ga^\mu q'_{jL})^2 [(V^*_{qL})_{1i}(V_{qL})_{1j}]^2\left(\fr{s^2_\varphi}{m^2_{Z_1}} + \fr{c^2_\varphi}{m^2_{Z_2}}\right).\ee The contribution of the standard model-like $Z_1$ boson is negligible too, since 
\be \fr{s^2_\varphi/m^2_{Z_1}}{c^2_\varphi/m^2_{Z_2}} \simeq \fr{[(1+\sqrt3\beta t^2_W)u^2 -(1-\sqrt3\beta t^2_W)v^2]^2}{4(u^2+v^2)w^2} < \left(\fr{1+\sqrt{3}t_W}{2}\right)^2 \fr{v^2_\mathrm{w}}{w^2}\simeq 0.95\fr{v^2_{\mathrm{w}}}{w^2},\ee which is suppressed due to $v_{\mathrm{w}}\ll w$. Above, we have used: i) $|\beta| <1/t_W$, which is derived from the conditions of the photon field normalization and the gauge coupling matching $s_W=e/g=t_X/\sqrt{1+(1+\beta^2)t^2_X}$ (partly aforementioned), and ii) $v_\mathrm{w}^2\equiv u^2+v^2=(246\ \mathrm{GeV})^2$, which is identified from the $W$ boson mass. It is easily proved that the $\rho$-parameter deviation from the standard model value due to the $Z$-$Z'$ mixing is obtained by $\Delta \rho=\rho-1\simeq (s^2_\varphi/m^2_{Z_1})/(c^2_\varphi/m^2_{Z_2})$. This again implies the nonsignificant contribution of $Z_1$ because of $\Delta\rho<0.0006$ from the global fit \cite{pdg}. Therefore, only the new field $Z_2$ governs the FCNCs, leading to 
\be \mathcal{L}^{\mathrm{eff}}_{\mathrm{FCNC}}\simeq \fr{1}{w^2} (\bar{q}'_{iL}\ga^\mu q'_{jL})^2 [(V^*_{qL})_{1i}(V_{qL})_{1j}]^2,\label{ttdd1}\ee which is independent of $\beta$ and the Landau pole, if this pole is presented for large $|\beta|$. This is a new observation of the present work, in agreement with a partial conclusion in \cite{s331}.     

In both economical 3-3-1 models discussed below, the ordinary ($u_a,d_a$) and exotic ($U,D_\al $) quarks that are correspondingly represented in the same triplet/antitriplet with the same electric charge might mix. Hence, the quark mixing matrices are redefined as $(u_1\ u_2\ u_3\ U)^T_{L,R}=V_{uL,uR}(u\ c\ t\ T)^T_{L,R}$ and $(d_1\ d_2\ d_3\ D_2\ D_3)^T_{L,R}=V_{dL,dR}(d\ s\ b\ B\ B')^T_{L,R}$, so that the $4\times 4$ mass matrix of up-type quarks ($u_a,U$) and the $5\times 5$ mass matrix of down-type quarks $(d_a,D_\al)$ are diagonalized \cite{dlfla}. The FCNC Lagrangian as coupled to $Z'$ is now changed to 
\be -\fr{g}{\sqrt{3(1-\beta^2 t^2_W)}}\bar{q}'_{iL}\ga^\mu q'_{jL}[V^\dagger_{qL}V_{qL}]_{ij}Z'_\mu,\ee where we denote $[V^\dagger_{uL}V_{uL}]_{ij}\equiv (V^*_{uL})_{1i}(V_{uL})_{1j}-\fr 1 2 (V^*_{uL})_{4i}(V_{uL})_{4j}$ for the up-type quarks and $[V^\dagger_{dL}V_{dL}]_{ij}\equiv (V^*_{dL})_{1i}(V_{dL})_{1j}+\fr 3 2 (V^*_{dL})_{4i}(V_{dL})_{4j}+\fr 3 2 (V^*_{dL})_{5i}(V_{dL})_{5j}$ for the down-type quarks. Correspondingly, the effective Lagrangian due to the $Z'$ contribution is achieved as 
\be \fr{1}{w^2} (\bar{q}'_{iL}\ga^\mu q'_{jL})^2[V^\dagger_{qL}V_{qL}]^2_{ij}.\label{ttdd2} \ee As mentioned in the above footnote, the ordinary and exotic quark mixings also lead to the FCNCs associated with $Z$, obtained by the Lagrangian, \be (\pm)\fr{g}{2c_W}\bar{q}'_{iL}\ga^\mu q'_{jL}(V^*_{qL})_{Ii}(V_{qL})_{Ij} Z_\mu,\ee where $``+"$ and $I=4$ are applied for $V_u$, whereas $``-"$ and $I=4,5$ are applied for $V_d$. Integrating $Z$ out, the corresponding effective Lagrangian is 
\be \fr{1}{v^2_{\mathrm{w}}} (\bar{q}'_{iL}\ga^\mu q'_{jL})^2[(V^*_{qL})_{Ii}(V_{qL})_{Ij}]^2.\label{ttdd3}\ee This contribution would spoil the standard model prediction for the neutral meson mass differences, if the mixing of the ordinary and exotic quarks was compatible to the ordinary quark mixing. For instance, the $K^0$-$\bar{K}^0$ mixing bounds $|(V^*_{dL})_{I1}(V_{dL})_{I2}|\lesssim 10^{-5}$, which is much smaller than the smallest CKM matrix element. To avoid the large FCNCs, we assume \be |(V^*_{qL})_{Ii}(V_{qL})_{Ij}| \ll |(V^*_{qL})_{1i}(V_{qL})_{1j}|,\label{ttddd1}\ee so that the $Z$ contribution (\ref{ttdd3}) is insignificant, and (\ref{ttdd2}) is thus reduced to (\ref{ttdd1}). The above inequality is also valid when the 1's are replaced by $\al=2,3$, due to the unitarity condition, $(V^\dagger_{qL} V_{qL})_{ij}=0$. Furthermore, the $B-L$ conservation demands that the exotic and ordinary quark mixings vanish \cite{3311,dhqt}. Hence, the suppressions like (\ref{ttddd1}) are naturally preserved by an approximate $B-L$ symmetry, as interpreted in \cite{d,dps,s331}. Lastly, there may exist tree-level FCNCs induced by the new non-Hermitian gauge bosons $X^{0,0*}=(A_4\mp i A_5)/\sqrt{2}$, which couple $u_1$ with $U$, and $d_\al $ with $D_\al $. The relevant Lagrangian is given by 
\bea \mathcal{L} &\supset& -\fr{g}{\sqrt{2}}(\bar{u}_{1L}\ga^\mu U_L - \bar{D}_{\al L}\ga^\mu d_{\al L}) X^0_\mu+H.c.\crn
&\supset& -\fr{g}{\sqrt{2}}[\bar{u}'_{iL}\ga^\mu u'_{jL} (V^*_{uL})_{1i}(V_{uL})_{4j}- \bar{d}'_{iL}\ga^\mu d'_{jL}(V^*_{dL})_{Ii}(V_{dL})_{\al j}] X^0_\mu+H.c., \eea where $I=2+\al$. This yields the effective Lagrangian,
\be \fr{1}{w^2}\left\{(\bar{u}'_{iL}\ga^\mu u'_{jL})^2[(V^*_{uL})_{4i}(V_{uL})_{1j}]^2+(\bar{d}'_{iL}\ga^\mu d'_{jL})^2[(V^*_{dL})_{Ii}(V_{dL})_{\al j}]^2\right\},\label{ttdd6}\ee where we have used $m^2_X=\fr{g^2}{4}(u^2+w^2)\simeq g^2w^2/4$. The $X$ boson contributions to the FCNCs (\ref{ttdd6}) are radically smaller than those of $Z'$ in (\ref{ttdd1}) due to the conditions (\ref{ttddd1}). In summary, for any 3-3-1 model the FCNCs due to $Z'$ in (\ref{ttdd1}) would dominate, which will be taken into account.       

Without loss of generality, by alignment in the up quark sector, i.e. $V_{uL}=1$, the CKM matrix is just $V_{\mathrm{CKM}}=V_{dL}$. The $K^0$-$\bar{K}^0$ mixing yields a bound \cite{pdg,mmbbdd},
\be \fr{1}{w^2} [(V^*_{dL})_{11}(V_{dL})_{12}]^2<\fr{1}{(10^4\ \mathrm{TeV})^2}.\ee
The CKM factor is $|(V^*_{dL})_{11}(V_{dL})_{12}|\simeq 0.22$ \cite{pdg}, which implies \be w>2.2\times 10^3\ \mathrm{TeV}. \label{bound1} \ee This high bound applies for the considering model with nonuniversal first quark generation. If one arranges the second quark generation differently from the others, the CKM factor is similarly $|(V^*_{dL})_{21}(V_{dL})_{22}|\simeq 0.22$ \cite{pdg}, which presents the same bound for $w$ as in the previous case. Furthermore, putting the third quark generation differently from the first two, the CKM factor is now smaller than the previous factors, i.e. $|(V^*_{dL})_{31}(V_{dL})_{32}|\simeq 3.5\times 10^{-4}$ \cite{pdg}, which yields \be w>3.5\ \mathrm{TeV}. \label{bound2}\ee 
Let us stress again that the bounds achieved in (\ref{bound1}) and (\ref{bound2}) are independent of $\beta$, applying for every 3-3-1 model with appropriate fermion content, i.e. quark arrangement. This is a new investigation of the present work, in agreement with the special cases in \cite{d,ds}.        

We can similarly study the bound for the $B^0_s$-$\bar{B}^0_s$ mixing, where $(i,j)=(2,3)$. One obtains $\fr{1}{w^2} [(V^*_{dL})_{12}(V_{dL})_{13}]^2<1/(100\ \mathrm{TeV})^2$ \cite{pdg,mmbbdd} for nonuniversal first quark generation, and so forth for other cases of quark arrangement. With the aid of the CKM factors in \cite{pdg}, if the second or third quark generation is arranged differently from the two others, it gives a bound $w>4$ TeV. Otherwise, when the first quark generation is differently arranged, it gives a negligible contribution to the $B$ meson mixing. We see that the $B$ mixing effect does not discriminate the second and third quark generations, unlike the case of the kaon mixing. The $B$ mixing gives the bound in agreement with the $K$ mixing when the third generation is differently arranged. However, it gives a negligible contribution to the $B$ mixing, when the kaon mixing bound is applied to the model with nonuniversal first or second quark generation.  

It is noteworthy that the bound (\ref{bound1}) applies for both the original economical 3-3-1 model \cite{e1,e2} and the reduced 3-3-1 model \cite{331rm}, where the first quark generation is nonuniversal. The latter model is ruled out as it is limited by a low Landau pole, $w\lesssim$ 5 TeV \cite{landaupole,s331}; additionally, it is encountered with a large $\rho$-parameter \cite{ds}. The former model presents a new physics at 1000 TeV scale. Of course, the previous predictions for the model at TeV are useless \cite{e3,e4,e5,e6,e7,e8,e9,e10,e11,ed1,ed2,ed3,ed4}. On the other hand, the bound (\ref{bound2}) is valid for both the minimal 3-3-1 model (including the simple 3-3-1 model as well) and the 3-3-1 model with right-handed neutrinos, where the third quark generation is nonuniversal, as often studied. We will also introduce a new economical 3-3-1 model working at TeV scale, avoiding the large bound (\ref{bound1}).

Let us remind the reader that the detailed outcomes of the FCNCs (\ref{ttdd1}) using the neutral meson mass differences are worth studying, but the overall bounds as obtained above would be expected (see, for instance, \cite{fcnc3,fcnc1}). In other words, it is sufficient for the purpose of this work as to classify and interpret the new directions of the economical 3-3-1 models, to be discussed below.        

\section{\label{t12e331} Two scenarios for the economical 3-3-1 model}

An economical 3-3-1 model is defined to work with the minimal fermion and scalar content that includes $\nu_{aR}$ in lepton triplets and only two scalar triplets, either $(\chi,\rho)$ or $(\chi,\eta)$. Such theory has an electric charge operator $Q=T_3-\fr{1}{\sqrt{3}}T_8+X$. As a result of the above analysis, there are two distinct economical 3-3-1 models. The first model has a particle content like the original economical 3-3-1 model (i.e., possessing nonuniversal first quark generation and $\chi,\rho$), but the 3-3-1 breaking scale is beyond 1000 TeV, called type-I economical 3-3-1 model. By contrast, the second model has nonuniversal third quark generation and $\chi,\eta$, which implies a TeV 3-3-1 breaking scale, called type-II economical 3-3-1 model.      

\subsection{Type-I economical 3-3-1 model}

The fermion and scalar content is given by \cite{e2,e4} 
\bea &&\psi_{aL} =  
\left(\begin{array}{c}
\nu_{aL}\\
e_{aL}\\
\nu^c_{aR}\end{array}\right)  \sim \left(1,3,-\fr{1}{3}\right),\hs e_{aR}\sim (1,1,-1),\\
 && Q_{1 L}= 
\left(\begin{array}{c}
u_{1L}\\
d_{1L}\\
U_{L}\end{array}\right)   \sim \left(3,3,\fr{1}{3}\right),\hs
 Q_{\al L} = 
\left(\begin{array}{c}
d_{\al L}\\
-u_{\al L}\\
D_{\al L} \end{array}\right)  \sim \left(3,3^*,0\right),\\ 
&&u_{a R}, U_R \sim \left(3,1,\fr 2 3\right),\hs d_{aR},D_{\al R}\sim \left(3,1,-\fr 1 3\right),\\  
&&\chi =
\left(
\begin{array}{c}
\chi^{0}_1\\
\chi^{-1}_2\\
\chi^{0}_3
\end{array}\right)\simeq \left(
\begin{array}{c}
\fr{1}{\sqrt{2}}u'+G^0_{X}\\
G^-_{Y}\\
\fr{1}{\sqrt{2}}(w+H_1+iG_{Z'}
\end{array}\right)\sim\left(1,3,-\fr{1}{3}\right),\\
&& \rho =
\left(
\begin{array}{c}
\rho^{+}_1\\
\rho^{0}_2\\
\rho^{+}_3
\end{array}\right)\simeq \left(
\begin{array}{c}
G^+_{W}\\
\fr{1}{\sqrt{2}}(v+H+iG_Z)\\
H^+_2
\end{array}\right)\sim \left(1,3,\fr{2}{3}\right).\eea Recall that $\al=2,3$, and the exotic quarks $U,D$ have ordinary electric charges, i.e. $Q(U)=2/3$ and $Q(D)=-1/3$ similar to $u$ and $d$, respectively. 

The 3-3-1 breaking scale is bounded by $w>2200$ TeV. Moreover, since $\chi^0_1$ has the lepton number $L=2\neq 0$, its VEV, $u'$, that breaks this charge should be much smaller than the weak scale $v$, i.e. $u'\ll v$. Indeed, because of $u'\neq 0$ there mix in the gauge boson sectors, the charged $W$-$Y$ and the neutral $Z$-$Z'$-$A_4$, in addition to the ordinary $Z$-$Z'$ mixing. Here $W^\pm=(A_1\mp i A_2)/\sqrt{2}$ and $Y^\mp=(A_6\mp i A_7)/\sqrt{2}$ denote the standard model and new gauge bosons, respectively, whereas all the other gauge fields including $X^{0,0*}$ have been already defined. Diagonalizing these sectors we get physical eigenstates and masses similarly to \cite{e2}. Consequently, from the $W$ boson mass, $m^2_W=g^2v^2/4$, we determine the weak scale $v\simeq 246$ GeV. The mixings in both the gauge boson sectors shift the tree-level $\rho$-parameter from the standard model prediction by $\Delta \rho=\rho -1 =\fr{m^2_W}{c^2_W m^2_Z}-1\simeq 3u'^2/v^2$, which implies $|u'|<3.5$ GeV due to the global fit $\Delta \rho<0.0006$~\cite{pdg}. Additionally, the elements of the mixing matrices between exotic and ordinary quarks are proportional to $u'/w\sim 10^{-6}$, which do not affect the FCNCs due to the $Z$ exchange as well as the non-unitarity of ordinary quark mixing matrices as remarked before \cite{dhqt}. 

Note that all the new particles, including the Higgs bosons $H_{1,2}$, the gauge bosons $Z',Y, X$, and the exotic quarks $U, D$, gain the masses proportional to the $w$ scale \cite{e2,e4}, which are very heavy, as expected. Furthermore, after the electroweak symmetry breaking, the ordinary particles ($W$, $Z$, $H$, charged leptons, and quarks) get consistent tree-level masses, expect for the following. As a property of the 3-3-1 model with minimal scalar content, there are 3 light quarks (one up and two down) that possess vanishing tree-level masses. However, they can obtain appropriate masses, induced by radiative corrections or effective interactions, according to the complete breakdown of the Peccei-Quinn symmetry, which was generally proved in \cite{e5,e11}.  

At the tree-level, the neutrinos have Dirac masses, one zero and two degenerate, which are unacceptable \cite{e6}. But, up to five-dimensional interactions, the relevant Yukawa Lagrangian is   
\be \mathcal{L}_{\mathrm{Yukawa}}\supset h^\nu_{ab}\bar{\psi}^c_{aL}\psi_{bL}\rho+\fr{h'^\nu}{\La}(\bar{\psi}^c_{aL}\psi_{bL})(\chi\chi)^*+H.c.,\ee where $\La$ is a cut-off scale, which can be taken as $\La\sim w$. Therefore, the observed neutrinos ($\sim \nu_L$) gain small Majorana masses via a seesaw mechanism, evaluated to be \be m_\nu\sim h^\nu (h'^\nu)^{-1} (h^\nu)^T \fr{v^2}{w},\ee which naturally fits the data since $w$ is as large as 2200 TeV. For instance, taking $m_\nu \sim 0.1$ eV and $h'^\nu\sim 1$ yields $h^\nu\sim 10^{-4}$, which is similar to the Yukawa couplings of the first- and second-generation fermions of the standard model. The heavy neutrinos ($\sim \nu_R$) have masses at $w$ scale. It is noted that the above neutrino mass generation scheme may be radiatively induced \cite{e6}.

The scalar field that breaks $SU(3)_L\otimes U(1)_X$ down to $SU(2)_L \otimes U(1)_Y$ is decomposed as $\chi^0_3=\fr{1}{\sqrt{2}}(w+H_1+iG_{Z'})$, where  $w$ provides the masses for all the new particles as well as setting the seesaw scale, as mentioned. Further, the imaginary part of this field is an unphysical Goldstone boson of $Z'$ that can be gauged away, while the real part includes a new, physical neutral Higgs boson, $H_1$, living at the $w$ scale. In the early universe, the full real field $\Phi=\sqrt{2}\Re({\chi^0_3})$ can be interpreted as an inflaton field involving (in time) toward the potential minimum $\Phi_{\mathrm{min}}=w$, driving the cosmic inflation. Let us consider the potential of $\Phi$ when the inflation scale is either not too high, but significantly larger than $w$, or close to the Planck scale.    

For the first case, the inflationary potential is radiatively contributed by the gauge bosons, the fermions, and the scalars, which couple to the inflaton. That said, it takes the form, \be V(\Phi)=\fr{\la}{4}(\Phi^2-w^2)^2+\fr{a}{64\pi^2}\Phi^4\ln \fr{\Phi^2}{w^2}+V_0,\ee up to the leading-log approximation \cite{cwp}. Here the renormalization scale has been fixed at $w$, and \be a\simeq \fr{13+4t^4_W}{48(3-t^2_W)^2} g^4 -\fr 1 2 (h^4_U+h^4_{D_2}+h^4_{D_3})+9\la^2+\fr 1 4 \la'^2.\ee The first term combines both the $SU(3)_L$ and $U(1)_X$ gauge boson contributions, with the substitution of $g_X=gt_W/\sqrt{1-t^2_W/3}$. Additionally, $h_{U,D_\al}$ denote the Yukawa couplings of the inflaton with exotic quarks $U,D_\al$, and $\la,\la'$ correspond to the self-inflaton and Higgs-inflaton quartic couplings, respectively. This potential yields an appropriate local minimum, given that $a/\la>-63.165$. Additionally, since $w$ is radically smaller than the inflation and Planck scales, i.e. $w\ll \Phi$, the inflationary potential is governed by the quartic and log terms. The number of e-folds will be chosen in the range $N\gtrsim 40$ so that the inflation scale is correspondingly higher than the expected 2200 TeV value. The cosmic microwave background (CMB) measurements yield a constraint on the curvature perturbation, which leads to $\la\lesssim 10^{-12}$ \cite{pdg}. Further, the spectral index $n_s$, the tensor-to-scalar ratio $r$, and the running index $\al$ can be evaluated as functions of $a'\equiv a/\la$ and are fitted to the experimental data \cite{pdg}. Then we obtain $a'\sim-10$, and thus $g\sim h_{U,D_\al }\sim \sqrt{\la,\la'}\lesssim 10^{-2.75}$, which contradicts the electroweak data $g\sim 0.5$. Conversely, this regime of the potential is not flat to reproduce a suitable inflation scenario.             

For the second case, the interaction of the inflaton to gravity via a non-minimal coupling $\xi$ may be important,
\be S=\int d^4x\sqrt{-g}\left[\fr 1 2(m^2_P + \xi \Phi^2)R+\fr 1 2 \pa^\mu \Phi \pa_\mu \Phi -\fr{\la}{4}(\Phi^2-w^2)^2 \right],\ee where $R$ is the scalar curvature, and $m_P=(8\pi G_N)^{-1/2}\simeq 2.4\times 10^{18}$ GeV is the reduced Planck mass. We assume $\xi\gtrsim 1$, and the action can be rewritten in the Einstein frame as \cite{higgsinflation}
\be S=\int d^4 x \sqrt{-\hat{g}}\left[\fr 1 2 m^2_P \hat{R} + \fr 1 2 \pa^\mu \phi \pa_\mu \phi -U(\phi)\right],\ee where the inflationary potential is related to the canonically-normalized inflaton field $\phi$ as
\be U(\phi)=\fr{\la m^4_P}{4\xi^2}\left(1+e^{-\sqrt{\fr{2}{3}}\fr{\phi}{m_P}}\right)^{-2}.\ee That said, the inflationary potential is flat due to the large field values, $\phi\gg m_P$ or $\Phi\gg m_P/\sqrt{\xi}$, and it successfully fits the data if $\xi\sim 10^4\sqrt{\la}$, in agreement to \cite{higgsinflation}. In this case, the number of e-folds set is about 60. Since $\la=m^2_{H_1}/(2w^2)$ can be small for a $H_1$ mass of a few TeV, the unitarity condition $\xi\lesssim \mathcal{O}(10)$ is recognized, and the inflation begins from the Planck regime $\Phi\sim m_P$. The reheating happens when the inflaton decays into the exotic quarks or the new gauge bosons. Considering the first case, it yields $T_R\sim h_{U,D_\al}(w/1000\ \mathrm{TeV})^{1/2}\times 10^{11}\ \mathrm{GeV}\sim 10^{11}$ GeV.        
 
Since the right-handed neutrinos do not directly couple to the inflaton, they could only be produced from the thermal bath of radiations. The CP-asymmetric decays of these right-handed neutrinos into a heavy charged Higgs boson and a charged lepton, $\nu_R\rightarrow H^\pm_2 e^\mp$, due to the Yukawa couplings $h^e_{ab}\bar{\psi}_{aL} \rho e_{bR}+H.c.$ can generate the expected baryon asymmetry via a leptogenesis mechanism similarly to the standard technique, provided that $m_{\nu_R}\gtrsim m_{H_2}$ \cite{leptogns}. However, it differs from the standard prediction due to the fact that the channels $\nu_R\rightarrow G^\pm_W e^\mp $ via the couplings $h^\nu_{ab}\bar{\psi}^c_{aL}\psi_{bL}\rho+H.c.$ are negligible, as suppressed by $h^\nu\ll h^\tau$ and $m_W\ll m_{H_2}$. Additionally, like the neutral field $H_1$, the finding of the charged field $H_2$ with some mass in the TeV regime can mark (suggest) the existence of this baryon-asymmetry production scheme.    

Let us emphasize that the economical 3-3-1 model has a natural room for dark matter as basic scalars filling up the model \cite{dps,s331}. As studied in \cite{dps}, the dark matter candidate might be resided in an inert scalar triplet, $\eta$, as a replication of $\chi$ under the gauge symmetry and an odd field under a $Z_2$ symmetry ($\eta\rightarrow -\eta$). We may have another inert scalar triplet responsible for dark matter as a replication of $\rho$ under the gauge symmetry, labeled $\rho'$, so that $\rho'\rightarrow-\rho'$ under a $Z_2$ symmetry. However, in the considering model, the candidate has a mass proportional to the 3-3-1 scale of order 1000 TeV. Therefore, if this mass is at or beyond this scale, the candidate cannot be generated as thermal relics as in \cite{dps}; otherwise, it overcloses the universe due to the unitarity constraint \cite{dmct}. Interestingly enough, this superheavy dark matter can be generated in the early universe by the mechanisms, such as gravitational and thermal productions, associated with the existing inflation and reheating, analogous to \cite{3311id}. By contrast, if the inert field masses are at TeV scale, the thermal generations may be interpreted as in \cite{dps}. 

Hence, by the realization of a high 3-3-1 breaking scale, the 3-3-1 model might simultaneously explain the neutrino masses and the cosmological issues, comparable to the other theories \cite{valle,d,ds,dhqt,3311,3311il,3311r,3311id}. Note that the usual 3-3-1 models do not reveal the inflation and associated superheavy dark matter. A detailed investigation of all the issues for this kind of the model is out of the scope of the present work, which should be published elsewhere \cite{dhsn}.                 
 
\subsection{Type-II economical 3-3-1 model}

A low bound for the 3-3-1 breaking scale is available only if the third quark generation is discriminative. In this case, the scalar triplet that breaks the electroweak symmetry should be $\eta$, instead of $\rho$, in order to generate the consistent top-quark mass (by contrast, if the scalar content as in the previous model is retained, the top quark has a vanishing tree-level mass that is impossible to be induced by subleading effects of radiative corrections or effective interactions, cf. \cite{s331,e5,e11} for details). Thus, the fermion and scalar content is appropriately derived as 
\bea &&\psi_{aL} =  
\left(\begin{array}{c}
\nu_{aL}\\
e_{aL}\\
\nu^c_{aR}\end{array}\right)  \sim \left(1,3,-\fr{1}{3}\right),\hs e_{aR}\sim (1,1,-1),\\
 &&  Q_{\al L} = 
\left(\begin{array}{c}
d_{\al L}\\
-u_{\al L}\\
D_{\al L} \end{array}\right)  \sim \left(3,3^*,0\right),\hs Q_{3 L}= 
\left(\begin{array}{c}
u_{3L}\\
d_{3L}\\
U_{L}\end{array}\right)   \sim \left(3,3,\fr{1}{3}\right),\\ 
&&u_{a R}, U_R \sim \left(3,1,\fr 2 3\right),\hs d_{aR},D_{\al R}\sim \left(3,1,-\fr 1 3\right),\\  
&&\chi =
\left(
\begin{array}{c}
\chi^{0}_1\\
\chi^{-1}_2\\
\chi^{0}_3
\end{array}\right)\simeq \left(
\begin{array}{c}
\fr{1}{\sqrt{2}}u'+c_\xi G^0_{X}-s_\xi H^0_2\\
G^-_{Y}\\
\fr{1}{\sqrt{2}}(w+H_1+iG_{Z'}
\end{array}\right)\sim\left(1,3,-\fr{1}{3}\right),\\
&& \eta =
\left(
\begin{array}{c}
\eta^{0}_1\\
\eta^{-1}_2\\
\eta^{0}_3
\end{array}\right)\simeq \left(
\begin{array}{c}
\fr{1}{\sqrt{2}}(u+H+iG_Z)\\
G^-_{W}\\
\fr{1}{\sqrt{2}}w'+s_\xi G^0_X+c_\xi H^0_2
\end{array}\right)\sim \left(1,3,-\fr{1}{3}\right),\eea where note that $\al=1,2$, $t_\xi=u'/w'$, and the physical scalar spectrum explicitly displayed can be obtained from the following scalar potential. Recall that the FCNC bounds yield: $w>3.5$ TeV for the $K$ mixing, and $w>4$ TeV for the $B_s$ mixing. Additionally, the flavor phenomenology of this kind of the 3-3-1 models has been extensively studied, for examples, in \cite{refre1,refre2,refre3}. 

The total Lagrangian is $\mathcal{L}=\mathcal{L}_{\mathrm{kinetic}}+\mathcal{L}_{\mathrm{Yukawa}}-V_{\mathrm{scalar}}$,
where 
\bea \mathcal{L}_{\mathrm{kinetic}} =\sum_F \bar{F}i\ga^\mu D_\mu F+\sum_S (D^\mu S)^\dagger(D_\mu S) -\fr 1 4 G_n^{\mu\nu}G_{n \mu\nu}-\fr 1 4 A_n^{\mu\nu}A_{n\mu\nu}-\fr 1 4 B^{\mu\nu}B_{\mu\nu}, \eea where $F,S$ run over fermion and scalar multiplets, respectively. $G_{n\mu\nu}$, $A_{n\mu\nu}$, and $B_{\mu\nu}$ are the field strength tensors corresponding to the 3-3-1 subgroups, respectively, and $D_\mu$ is the covariant derivative previously supplied. The Yukawa Lagrangian and scalar potential are 
\bea \mathcal{L}_{\mathrm{Yukawa}}&=&h^U_{33}\bar{Q}_{3L}\chi
U_{R}+h^D_{\al\beta}\bar{Q}_{\al L}\chi^* D_{\beta R} +h^u_{3a}
\bar{Q}_{3L}\eta u_{aR}+h^d_{\al a}\bar{Q}_{\al L}\eta^* d_{aR}\crn
&& + h'^u_{3a}\bar{Q}_{3L}\chi u_{aR}+h'^d_{\al
a}\bar{Q}_{\al L}\chi^* d_{a R}+h'^U_{33}\bar{Q}_{3L}\eta U_R+h'^D_{\al
\bet}\bar{Q}_{\al L}\eta^* D_{\bet R}+ H.c.,\label{yuka}\\
 V_{\mathrm{scalar}}&=&
\mu^2_1\eta^\dagger\eta+\mu_2^2\chi^\dagger\chi+\la_1(\eta^\dagger\eta)^2+\la_2(\chi^\dagger\chi)^2+\la_3(\eta^\dagger\eta)
(\chi^\dagger\chi)+\la_{4}(\eta^\dagger\chi)
(\chi^\dagger\eta)\crn
&&+\left[\mu'^2_3\eta^\dagger
\chi+\la'_5(\eta^\dagger \chi)^2+
(\la'_{6}\eta^\dagger\eta
+\la'_{7}\chi^\dagger\chi)\eta^\dagger\chi +H.c.\right].\label{smppp}\eea 

As established in \cite{3311,d}, in the general 3-3-1 model, the baryon minus lepton number $B-L$ neither commutes nor closes algebraically with $SU(3)_L$. For instance, with $L(\nu_R)=1$ and $B(\nu_R)=0$, a lepton triplet has $B-L=\mathrm{diag}(-1,-1,1)$, which does not commute with the $T_{4,5,6,7}$ generators of $SU(3)_L$. Additionally, if the algebras are closed, $B-L$ must be some generator of $SU(3)_L$, $B-L=x_n T_n$, which yields $\mathrm{Tr}(B-L)=0$, in contrast with the lepton triplet $\mathrm{Tr}(B-L)=-1$. Indeed, it is clear that the minimal interactions of the model (the unprimed couplings) conserve a new Abelian symmetry, $U(1)_N$, that along with $SU(3)_L$ close those symmetries, realizing $B-L=-\fr{4}{\sqrt{3}}T_8+N$ as a residual charge of $SU(3)_L\otimes U(1)_N$. The charges $N$ and $X$ are independent as $B-L$ and $Q$ are. The $N$-charges for the multiplets are obtained as $N(\psi_{aL},Q_{3L},Q_{\al L},e_{aR},u_{aR},d_{aR},U_R,D_{\al R},\eta,\chi)=-1/3,1,-1/3,-1,1/3,1/3,7/3,-5/3,2/3,-4/3$, respectively. Moreover, the nontrivial $B-L$ charges for new particles are collected as $[B-L](U,D,\eta^0_3,\chi^0_1,\chi^-_2,X^0,Y^-)=7/3,-5/3,2,-2,-2,-2,-2$, respectively. Here, the fields $X$ and $Y$ are the non-Hermitian gauge bosons respectively coupled to $T_{4,5}$ and $T_{6,7}$, as mentioned. 

It is easily checked that the nonminimal Yukawa couplings, those primed in (\ref{yuka}), violate $B-L$ by two units, while the nonminimal scalar-couplings and mass-parameters, those primed in (\ref{smppp}), violate this charge by one or two units, respectively. Furthermore, since the scalar fields $\eta^0_3$ and $\chi^0_1$ have $B-L\neq 0$, their VEVs $u',w'$ break $B-L$. This is in contrast with the normal VEVs $u,w$, which carry no $B-L$ and conserve this charge. Additionally, all the above ingredients are necessarily included to realize $B-L$ as an approximate symmetry; otherwise, the 3-3-1 model is not self-consistent, warranting a 3-3-1-1 gauge extension \cite{d}. For consistency with the standard model, the violating parameters such as the couplings and the VEVs should be much smaller than the corresponding conserved ones, $u'\ll u,\ w'\ll w,\ h'\ll h,\ \la'\ll \la$, etc. Additionally, $u\simeq 246\ \mathrm{GeV}$ is extracted from the $W$ boson mass, which implies $u \ll w$.        

It is easily justified that the leptons and three ordinary quarks (two up quarks and one down quark) have vanishing tree-level masses. Furthermore, the Lagrangian of the model automatically contains (i.e., conserves) the Peccei-Quinn--like symmetries, similarly to the original economical 3-3-1 model \cite{e11}. Such massless particles can get appropriate masses when the Peccei-Quinn--like symmetries are completely broken via radiative corrections or effective interactions \cite{e11}. Let us impose the latter which is given, up to five dimensions, by     
\bea  \mathcal{L}'_{\mathrm{Yukawa}} &=& \fr{1}{\La}(\bar{Q}_{3L} \eta^*\chi^*)(h^d_{3a} d_{aR}+h'^D_{3\al} D_{\al R}) + \fr{1}{\La}(\bar{Q}_{\al L}\eta \chi) (h^u_{\al a} u_{aR}+h'^U_{\al 3} U_{R})\crn
&&+\fr{1}{\La}h^e_{ab}\bar{\psi}_{aL}\eta^*\chi^* e_{bR} + \fr{1}{\La}(\bar{\psi}^c_{aL} \psi_{bL} ) (f'^\nu_{ab}\eta\eta+g'^\nu_{ab}\chi\chi+h'^\nu_{ab}\eta\chi)^* + H.c.,\label{yeff}\eea where the unprimed couplings conserve $B-L$, while the primed couplings stand for the violating ones, as usual. Additionally, the quark and neutrino effective couplings explicitly violate the Peccei-Quinn-like charges \cite{e11}. The cutoff scale $\La$ can be taken in the same order as $w$. Specially $f'^\nu_{ab}$ and $g'^\nu_{ab}$ are symmetric in flavor indices, whereas $h'^\nu_{ab}$ is a generic matrix. 

Substituting the VEVs of the scalars into the relevant Lagrangians in (\ref{yuka}) and (\ref{yeff}), all the fermion mass matrices are derived. Using the conditions $u'\ll u$ and $w'\ll w$ hereafter, the charged leptons obtain masses, $[M_e]_{ab}\simeq h^e_{ab} u \fr{w}{2\La}$. Since $w\sim \La$ and $u\simeq 246$ GeV, the $M_e$ fits the measured masses of charged leptons, analogous to the standard model. We have a seesaw mechanism for the neutrino masses, which works due to $u\ll w\sim \La$. Indeed, the right-handed neutrinos achieve large Majorana masses, $[M_{R}]_{ab}\simeq -g'^\nu_{ab}\fr{w^2}{\La}$. The left-handed neutrinos gain small Majorana masses, $[M_L]_{ab}\simeq -f'^\nu_{ab}\fr{u^2}{\La}$. The neutrino Dirac masses take the form, $[M_D]_{ab}\simeq -h'^\nu_{ab}\fr{uw}{2\La}$. Thus, the observed neutrinos $(\sim \nu_L)$ obtain small masses via a combined type I and II seesaw mechanism, \be M_\nu\simeq M_L-M^T_D M^{-1}_R M_D\simeq -\fr{u^2}{\La}\left[f'^\nu-\fr 1 4 (h'^\nu)^T(g'^\nu)^{-1}h'^\nu\right].\ee 

The new observation is that the neutrinos get masses when both the Peccei-Quinn--like and $B-L$ symmetries are broken. The strength of the symmetry breakings is set by the primed couplings of the effective interactions, commonly called $h'$, thus $M_\nu\sim \fr{u^2}{\La}h'$. Note that for the 3-3-1 model, if $B-L$ is conserved, it must be a gauged charge, and that the effective interactions (primed) must be absent \cite{d,ds,3311,dhqt,3311il,3311id,3311r}. Therefore, $h'$ measures the approximate $B-L$ symmetry as well as the nonunitarity of the 3-3-1 model, as imprinted from the 3-3-1-1 model. The $h'$ strength can be obtained by integrating the $U(1)_N$ gauge boson out from the 3-3-1-1 model, which matches $h'/\La=g_N/\La_{N}$. Further, we have $h'\sim \La/\La_N\sim 10^{-11}$, where $\La_N\sim 10^{14}$ GeV is just the inflation scale and $g_N\sim 1$ \cite{3311il,3311id}. This implies $M_\nu\sim 0.1$ eV as desirable. Alternatively, comparing $M_\nu/M_e\sim \fr{u}{w}\fr{h'}{h}$ with $u/w\sim0.1$ and $M_\nu/M_e\sim 10^{-6}$, it yields $h'/h\sim 10^{-5}$. Thus, the breaking strength $h'$ is suitably smaller than the electron Yukawa coupling, in agreement to \cite{s331}.              

At this stage, an evaluation shows that all the ordinary quarks obtain consistent masses, in agreement to \cite{e11}. Moreover, the elements of the mixing matrices of the exotic and ordinary quarks are proportional to $u'/u$, $w'/w$, and $h'/h$---the ratios of the $B-L$ violating parameters over the corresponding normal ones \cite{dlfla}. Again, the VEVs $u',w'$ and the couplings $h'$ should be small, $u'\ll u,\ w'\ll w,\ h'\ll h$, in order to suppress the dangerous FCNCs coming from $Z$ boson exchange due to the ordinary and exotic quark mixings. Generalizing the above result as well as in \cite{dhqt}, we obtain $u'/u\sim w'/w\sim h'/h\sim \sqrt{|(V^*_{dL})_{I1}(V_{dL})_{I2}|}\lesssim 3.16\times 10^{-3}$, where $(V_{dL})_{Ii}$ is the element that correspondingly connects the exotic and ordinary quarks in the mixing matrix. It yields $u'\lesssim 0.77$ GeV due to $u=246$ GeV, and $w'\lesssim 3.16$, $15.8$, and $31.6$ GeV for $w=1$, 5, and 10 TeV, respectively. Also, $h'$ for the quark sector is more suppressed, similarly to the ones for the neutrino masses. In practice, the VEVs $u',w'$ break $B-L$ (i.e., the lepton number), and that they are suppressed to be small due to the corresponding lepton-number violating scalar-potential. From the conditions of the potential minimization, we have roundly $u'\sim\la'_7u$ and $w'\sim\la'_7w$. Thus, $u'$ and $w'$ should be small since its absence, i.e. $\la'_7=0$, enhances the 3-3-1-1 gauge symmetry.            

Following the approach in \cite{dps,s331}, the model can provide realistic dark matter candidates. If one introduces the inert triplet $\rho$---which is analogous to the field in the 3-3-1 model with right-handed neutrinos but is odd under a $Z_2$ symmetry---it cannot be dark matter. Indeed, the candidate $\rho^0_2=\fr{1}{\sqrt{2}}(H+iA)$ resided in $\rho$ yields degenerate masses for $H$ and $A$, which implies a large direct detection cross-section via $Z$ exchange. This is already ruled out by the experiment \cite{dmexp}. However, an inert triplet as replication of $\eta$ or $\chi$ under the gauge symmetry, called $\zeta=(\zeta^0_1,\zeta^-_2,\zeta^0_3)$, that transforms nontrivially under a $Z_2$ symmetry ($\zeta\rightarrow -\zeta$) might provide a consistent candidate as the combination of either real or imaginary parts of $\zeta^0_{1,3}$. The inert scalar sextet responsible for dark matter can be also interpreted, similarly to the simple 3-3-1 model \cite{s331}. The details of the dark matter identification and stability proof could be similarly achieved as in \cite{s331,dps}, which are not further discussed. That said, the model predicts those candidates as WIMPs at TeV scale. 

In summary, the 3-3-1 model with right-handed neutrinos has a nontrivial vacuum for $u'\neq 0$ and $w'\neq 0$, and this yields the appropriate new-physics consequences as obtained. Interestingly, the type II economical 3-3-1 model is a minimal realization of this vacuum, while it explicitly indicates to dark matter. See \cite{radn331} for other interpretations. Note that the previous studies \cite{331r} only consider the vacuum with $u'=w'=0$, and thus the above consequences were not recognized although they include more than two scalar triplets.                        
 
\section{\label{con} Conclusion}

As a fundamental property, the 3-3-1 model presents the FCNCs associated with $Z'$ boson due to nonuniversal fermion generations under the gauge symmetry. We have proved that the FCNCs that describe neutral meson mixings are independent of both the embedding of electric charge operator and the potential Landau pole. Applying the result for the $K$ and $B_s$ mixings, we obtain the new physics scale: (a) $w>2200$ TeV, if the first or second fermion generation is discriminative, and (b) $w>3.5$ TeV for  the $K$ system and $w>4$ TeV for the $B_s$ system, if the third fermion generation is discriminative. 

Due to the above constraint, the original economical 3-3-1 model (named type-I) works in a large energy regime of order 1000 TeV, yielding simultaneously the novel consequences of the neutrino mass generation scheme, cosmic inflation, leptogenesis, and superheavy dark matter. The 3-3-1 breaking field, $\chi^0_3$, is important to set the seesaw scale $w$, which originates from the inflation scale, and define the inflaton $\Phi$. The decays of $\Phi$ to pairs of new quarks or of new gauge bosons reheat the universe. The CP-violating decays of $\nu_R$ to a heavy charged Higgs ($H^\pm_2$) and charged lepton govern the baryon asymmetry. Dark matter is a hidden/inert scalar field, a replication of $\chi$ (called $\eta$) or a replication of $\rho$ (called $\rho'$), which might be created in the early universe by nonthermal processes/mechanisms associated with the inflation and reheating. Alternatively, the light candidates may play the role of WIMPs. The imprints of the inflation and leptogenesis mechanisms at the TeV scale are just the new Higgs fields $H_{1,2}$, which may be verified at the LHC.     

Alternatively, we have introduced a new economical 3-3-1 model (called type-II), where the third fermion generation is rearranged differently from the first two generations, and that the scalar content includes $\eta,\chi$. This model works naturally at the TeV scale, providing interesting results. The lepton number breaking/violating parameters are suppressed, $u'\ll u$, $w'\ll w$, $h'\ll h$, and $\la'\ll \la$, by the approximate $B-L$ symmetry. The strength of the lepton number breaking might have a source from the 3-3-1-1 breaking to be naturally small, responsible for the neutrino masses. Moreover, the approximate $B-L$ symmetry strictly prevents the dangerous FCNCs coming from the ordinary and exotic quark mixings, bounding the violating parameters to be $u',w'\sim \mathcal{O}(1)$ GeV and $h'/h\lesssim3.16\times 10^{-3}$ for the quark couplings. Both the neutrinos and quarks gain consistent masses also associated with the complete breakdown of the Peccei-Quinn--like symmetries. It is shown that a hidden scalar field $\zeta$ as a replication of $\eta$ or $\chi$ can provide appropriate WIMP thermal relics. However, if $\rho$ is included as an inert scalar, it cannot be dark matter. 

Let us stress that the discrimination of fermion generations as recognized at a scale of order 1000 TeV is surprisingly close to the WIMP mass limit $\sim 500$ GeV \cite{dmct}. Although the 3-3-1 model does not directly solve this coincidence, it provides both the scenarios for dark matter as the nonthermal and thermal relics. Therefore, these two economical 3-3-1 models would predict and connect the particle physics to cosmological issues with rich phenomenologies, attracting much attention \cite{dhsn}.                  

\section*{Acknowledgments}

This research is funded by Vietnam National Foundation for Science and Technology Development (NAFOSTED) under grant number 103.01-2016.44.


\begin{thebibliography}{99}

\bibitem{pdg} C. Patrignani {\it et al.} (Particle Data Group), Chin. Phys. C {\bf 40}, 100001 (2016).

\bibitem{331m} F. Pisano and V. Pleitez, Phys. Rev.  D {\bf 46}, 410 (1992);
P. H. Frampton, Phys. Rev. Lett. {\bf 69}, 2889 (1992); R. Foot,
O. F. Hernandez, F. Pisano, and V. Pleitez, Phys. Rev. D {\bf 47},
4158 (1993).

\bibitem{331r} M. Singer, J. W. F. Valle, and J. Schechter, Phys.
Rev. D {\bf 22}, 738 (1980); J. C. Montero, F. Pisano, and V.
Pleitez, Phys. Rev. D {\bf 47}, 2918 (1993); R. Foot, H. N. Long,
and Tuan A. Tran, Phys. Rev. D {\bf 50}, R34 (1994).       

\bibitem{d} P. V. Dong, Phys. Rev. D {\bf 92}, 055026 (2015).

\bibitem{331beta1} R. A. Diaz, R. Martinez, and F. Ochoa, Phys. Rev. D {\bf 69}, 095009 (2004).  

\bibitem{331beta2} R. A. Diaz, R. Martinez, and F. Ochoa Phys. Rev. D {\bf 72}, 035018 (2005). 

\bibitem{e1} W. A. Ponce, Y. Giraldo, and L. A. Sanchez, Phys. Rev. D {\bf 67},
075001 (2003). 

\bibitem{e2} P. V. Dong, H. N. Long, D. T. Nhung, and D. V. Soa, Phys. Rev. D {\bf 73}, 035004 (2006). 

\bibitem{e3} P. V. Dong, H. N. Long, and D. T. Nhung, Phys. Lett. B {\bf 639}, 527 (2006). 

\bibitem{e4} P. V. Dong, H. N. Long, and D. V. Soa, Phys. Rev. D {\bf 73}, 075005 (2006). 

\bibitem{e5} P. V. Dong, Tr. T. Huong, D. T. Huong, and H. N. Long, Phys. Rev. D {\bf 74}, 053003 (2006). 

\bibitem{e6} P. V. Dong, H. N. Long, and D. V. Soa, Phys. Rev. D {\bf 75}, 073006 (2007).  

\bibitem{e7} P. V. Dong, D. T. Huong, M. C. Rodriguez, and H. N. Long, Nucl. Phys. B {\bf 772}, 150 (2007). 

\bibitem{e8} P. V. Dong, Tr. T. Huong, N. T. Thuy, and H. N. Long, JHEP {\bf 11}, 073 (2007). 

\bibitem{e9} P. V. Dong, D. T. Huong, N. T. Thuy, and H. N. Long, Nucl. Phys. B {\bf 795}, 361 (2008). 

\bibitem{e10} D. V. Soa, P. V. Dong, Tr. T. Huong, and H. N. Long, J. Exp. Theor. Phys. {\bf 108}, 757 (2009). 

\bibitem{e11} P. V. Dong, H. N. Long, and H. T. Hung, Phys. Rev. D {\bf 86}, 033002 (2012).

\bibitem{ed4} Y. Giraldo, W. A. Ponce, and L. A. S\'anchez, Eur. Phys. J. C {\bf 63}, 461 (2009).  

\bibitem{ed1} R. H. Benavides, L. N. Epele, H. Fanchiotti, C. G. Canal, and W. A. Ponce, Adv. High Energy Phys. {\bf 2015}, 813129 (2015). 

\bibitem{ed2} J. C. Montero and B. L. S\'anchez-Vega, Phys. Rev. D {\bf 84}, 055019 (2011); {\bf 91}, 037302 (2015).

\bibitem{ed3} V. Q. Phong, H. N. Long, V. T. Van, and L. H. Minh, Eur. Phys. J. C {\bf 75}, 342 (2015).   

\bibitem{e12} D. Cogollo, arXiv:1706.00397 [hep-ph].

\bibitem{331rm} J. G. Ferreira Jr, P. R. D. Pinheiro, C. A. de S. Pires, and P. S. Rodrigues da Silva, Phys. Rev. D {\bf 84}, 095019 (2011).  

\bibitem{ds} P. V. Dong and D. T. Si, Phys. Rev. D {\bf 90}, 117703 (2014). 

\bibitem{s331} P. V. Dong, N. T. K. Ngan, and D. V. Soa, Phys. Rev. D {\bf 90}, 075019 (2014). 

\bibitem{s331dm} P. V. Dong, C. S. Kim, N. T. Thuy, and D. V. Soa, Phys. Rev. D {\bf 91}, 115019 (2015).

\bibitem{s331pheno} P. V. Dong and N. T. K. Ngan, arXiv:1512.09073 [hep-ph].  

\bibitem{landaupole} A. G. Dias, R. Martinez, and V. Pleitez, Eur. Phys. J. C {\bf 39}, 101 (2005); A.G. Dias, Phys. Rev. D {\bf 71}, 015009
(2005).

\bibitem{dhqt} P. V. Dong, D. T. Huong, F. S. Queiroz, and N. T. Thuy, Phys. Rev. D {\bf 90}, 075021 (2014). 

\bibitem{dl} P. V. Dong and H. N. Long, Eur. Phys. J. C {\bf 42}, 325 (2005).

\bibitem{fcnc3} D. Cogollo, A. Vital de Andrade, F. S. Queiroz, and P. R. Teles, Eur. Phys. J. C {\bf 72}, 2029 (2012). 

\bibitem{fcnc1} A. C. B. Machado, J. C. Montero, and V. Pleitez, Phys. Rev. D {\bf 88}, 113002 (2013). 

\bibitem{fcnc2} H. Okada, N. Okada, Y. Orikasa, and K. Yagyu, Phys. Rev. D {\bf 94}, 015002 (2016).

\bibitem{dlfla} P. V. Dong and H. N. Long, Phys. Rev. D {\bf 77}, 057302 (2008) [arXiv:0801.4196v1 [hep-ph]]. 

\bibitem{3311} P. V. Dong, T. D. Tham, and H. T. Hung, Phys. Rev. D {\bf 87}, 115003 (2013). 

\bibitem{dps} P. V. Dong, T. Phong Nguyen, and D. V. Soa, Phys. Rev. D {\bf 88}, 095014 (2013).

\bibitem{mmbbdd} M. Bona {\it et al.} (UTfit Collaboration), JHEP {\bf 03}, 049 (2008); G. Isidori, Y. Nir, and G. Perez, Ann. Rev. Nucl. and Part. Sci. {\bf 60}, 355 (2010). 

\bibitem{cwp} S. R. Coleman and E. Weinberg, Phys. Rev. D {\bf 7}, 1888 (1973).

\bibitem{higgsinflation} F. L. Bezrukov and M. Shaposhnikov, Phys. Lett. B {\bf 659}, 703 (2008). 

\bibitem{leptogns} M. Fukugita and T. Yanagida, Phys. Lett. B {\bf 174}, 45 (1986). 

\bibitem{dmct} K. Griest and M. Kamionkowski, Phys. Rev. Lett. {\bf 64}, 615 (1990). 

\bibitem{3311id} D. T. Huong and P. V. Dong, Eur. Phys. J. C {\bf 77}, 204 (2017).  

\bibitem{3311il} D. T. Huong, P. V. Dong, C. S. Kim, and N. T. Thuy, Phys. Rev. D {\bf 91}, 055023 (2015). 

\bibitem{3311r} A. Alves, G. Arcadi, P. V. Dong, L. Duarte, F. S. Queiroz, and J. W. F. Valle, Phys. Lett. B {\bf 772}, 825 (2017) [arXiv:1612.04383 [hep-ph]]. 

\bibitem{valle} S. M. Boucenna, S. Morisi, Q. Shafi, and J. W. F. Valle, Phys. Rev. D {\bf 90}, 055023 (2014).

\bibitem{dhsn} P. V. Dong, D. T. Huong, T. T. Nhat, and D. V. Soa, ``3-3-1 model implication for neutrino masses and cosmological issues'', in preparation.

\bibitem{refre1} Andrzej J. Buras and Fulvia De Fazio, JHEP {\bf 08}, 115 (2016) [arXiv:1604.02344 [hep-ph]].  

\bibitem{refre2} Andrzej J. Buras, Fulvia De Fazio, Jennifer Girrbach, and Maria V. Carlucci, JHEP {\bf 02}, 023 (2013) [arXiv:1211.1237 [hep-ph]].

\bibitem{refre3} Christoph Promberger, Sebastian Schatt, and Felix Schwab, Phys. Rev. D {\bf 75}, 115007 (2007) [arXiv:hep-ph/0702169].

\bibitem{dmexp} R. Barbieri, L. J. Hall, and V. S. Rychkov, Phys. Rev. D {\bf 74}, 015007 (2006). 

\bibitem{radn331} S. M. Boucenna, S. Morisi, and J. W. F. Valle, Phys. Rev. D {\bf 90}, 013005 (2014); S. M. Boucenna, R. M. Fonseca, F. Gonzalez-Canales, and J. W. F. Valle, Phys. Rev. D {\bf 91}, 031702 (2015); J. W. F. Valle and C. A. Vaquera-Araujo, Phys. Lett. B {\bf 755}, 363 (2016). 
     

\end{thebibliography}
 \end{document}